\documentclass[twocolumn,showpacs,preprintnumbers,amsmath,amssymb,superscriptaddress,longbibliography,floatfix]{revtex4-2}
\usepackage[utf8]{inputenc}
\usepackage{amsmath}
\usepackage{upgreek}
\usepackage{graphicx}
\usepackage{xr}
\usepackage{color}
\usepackage{float}
\usepackage{multirow}

\makeatletter
\newcommand*{\addFileDependency}[1]{% argument=file name and extension
  \typeout{(#1)}
  \@addtofilelist{#1}
  \IfFileExists{#1}{}{\typeout{No file #1.}}
}
\makeatother
 
\newcommand*{\myexternaldocument}[1]{%
    \externaldocument{#1}%
    \addFileDependency{#1.tex}%
    \addFileDependency{#1.aux}%
}

\myexternaldocument{supplemental_material}
\begin{document}

\title{Quantitative diffractometric biosensing}% Force line breaks with \\

\author{Yves Blickenstorfer}
\altaffiliation{%
	A. Frutiger and Y. Blickenstorfer contributed equally to this work.
}%
\affiliation{%
	Laboratory of Biosensors and Bioelectronics, Institute of Biomedical Engineering, ETH Zürich, 8092 Zürich, Switzerland
}%
\author{Markus Müller}
\affiliation{%
Condensed Matter Theory Group, Paul Scherrer Institute, 5232 Villigen, Switzerland
}%
\author{Roland Dreyfus}
\affiliation{%
	Laboratory of Biosensors and Bioelectronics, Institute of Biomedical Engineering, ETH Zürich, 8092 Zürich, Switzerland
}%
\author{Andreas Michael Reichmuth}
\affiliation{%
Laboratory of Biosensors and Bioelectronics, Institute of Biomedical Engineering, ETH Zürich, 8092 Zürich, Switzerland
}%

\author{Christof Fattinger}
\email{christof.fattinger@gmail.com}
\affiliation{
	Roche Pharma Research and Early Development, Roche Innovation Center Basel, 4070 Basel, Switzerland
}%

\author{Andreas Frutiger}
\email{afrutig@ethz.ch}
\affiliation{%
	Laboratory of Biosensors and Bioelectronics, Institute of Biomedical Engineering, ETH Zürich, 8092 Zürich, Switzerland
}%

\date{\today}% It is always \today, today,
  % but any date may be explicitly specified

\begin{abstract}
Diffractometric biosensing is a promising technology to overcome critical limitations of refractometric biosensors, the dominant class of label-free optical transducers. These limitations manifest themselves by higher noise and drifts due to insufficient rejection of refractive index fluctuations caused by variation in temperature, solvent concentration, and most prominently, non-specific binding. Diffractometric biosensors overcome these limitations with inherent self-referencing on the submicron scale with no compromise on resolution. Despite this highly promising attribute, the field of diffractometric biosensors has only received limited recognition. A major reason is the lack of a general quantitative analysis. This hinders comparison to other techniques and amongst different diffractometric biosensors. For refractometric biosensors, on the other hand, such a comparison is possible by means of the refractive index unit (RIU). In this publication, we suggest the coherent surface mass density, $\Gamma_{\rm{coh}}$, as a quantity for label-free diffractometric biosensors with the same purpose as RIU in refractometric sensors. It is easy to translate $\Gamma_{\rm{coh}}$  to the total surface mass density $\Gamma_{\rm{tot}}$, which is an important parameter for many assays. We provide a generalized framework to determine $\Gamma_{\rm{coh}}$ for various diffractometric biosensing arrangements which enables quantitative comparison. Additionally, the formalism can be used to estimate background scattering in order to further optimize sensor configurations. Finally, a practical guide with important experimental considerations is given to enable readers of any background to apply the theory. Therefore, this paper provides a powerful tool for the development of diffractometric biosensors and will help the field to mature and unveil its full potential.

\end{abstract}

\maketitle

\section*{Introduction}

Optical transducers represent one of the predominant classes in the field of biomolecular sensing with numerous applications ranging from clinical diagnostics to drug discovery \cite{Turner2013oz}. For diagnostics, labeled technologies based on fluorescence or nanoparticle scattering are prevailing due to their high sensitivity \cite{Hosseini2018-zq, Sajid2015-vm}. In drug discovery, it is important to study the interaction kinetics between biochemical components, which is why label-free approaches are employed. This field is dominated by refractometric sensors \cite{Cooper2011-qy}. The refractometric sensing principle works as follows: A surface is modified by an adlayer containing immobilized receptors. The receptors interact specifically with target molecules. Binding of target molecules to the adlayer results in a change of the real part of the refractive index, which is then measured by the transducer. The most prominent refractometric transducer technology is surface plasmon resonance (SPR) \cite{Schasfoort2017-hs}. Its outstanding sensitivity can be illustrated by the fact that it can resolve less than one protein molecule per $\upmu$m$^2$ \cite{Homola2008-px}. Nevertheless, refractometric sensors are limited by the fact that any fluctuation of the refractive index in the entire volume of the evanescent field alters the signal \cite{Frutiger2019-um}. Such fluctuations can be caused by temperature variations, changes in sample composition and non-specific binding. diffractometric biosensing (also known as diffraction based sensing or diffractive optics technology) is a promising technology to overcome these drawbacks as will be explained later \cite{Gatterdam2017-bk}.

First, we focus on the physical principle behind diffractometric biosensing. Diffraction describes the phenomena of a wave bending around obstacles. In optics, such obstacles can be described by the difference in refractive index ($n+i\kappa$) compared to the refractive index of the propagation medium. This can be a difference in the real (phase) or imaginary (absorption) part of the refractive index. For significant diffraction, the size of the obstacle must be of the same order of magnitude as the wavelength. By arranging sub-wavelength obstacles in a spatially regular, \textit{i.e.} coherent pattern the diffraction from the pattern can be precisely tailored and enhanced in specific directions. For example, a periodic pattern diffracts coherent light of a given wavelength into specific directions that are determined by the diffraction order. Such coherent patterns are commonly used in the form of diffraction gratings. As described above, optical diffraction can be due to a disturbance of phase or amplitude. Phase gratings, which modulate the refractive index $n$, find various applications such as in spectrometers \cite{Loewen2018-tb}, as coupling elements in integrated optical chips \cite{Hunsperger2009-zp}, as phasemasks in photolithography \cite{Gatterdam2017-bk} and even for refractometric sensing \cite{Tiefenthaler1989-cj}. An important parameter to asses the performance of gratings is the diffraction efficiency. It is the ratio between the power diffracted into a certain order and the incident power, which is relatively easy to measure experimentally. However, its mathematical description for a given diffractive set-up is complicated and no generally applicable formulas exist \cite{Loewen2018-tb}. Nevertheless, it holds generally that for weak phase gratings the diffraction efficiency increases with the area, the thickness and the refractive index contrast of a diffractive structure \cite{Loewen2018-tb}.

After these general considerations we return to the specifics of diffractometric biosensing. In all implementations, the sensor consists of a diffraction grating. Binding of the target analyte changes the properties of the grating such as the grating thickness or the refractive index contrast. This results in a measurable change of the diffraction efficiency. The concept should not be confused with refractometric sensing based on diffraction gratings, where the spectral response (such as the direction of a certain diffraction order) is measured due to a change in the refractive index surrounding the grating \cite{Tiefenthaler1989-cj}. Most label-free diffractometric sensors make use of the same bio-physical property as refractometric sensors: the relatively high refractive index of biomolecules compared to that of water \cite{McMEEKIN1964-er}. To measure this refractive index sensitively, receptors are arranged in a coherent pattern so as to form a diffraction grating. Binding of target molecules to the pattern alters the refractive index contrast of the grating and therefore increases the diffraction efficiency. Ideally, the structure of receptors without bound target molecules has a very low contrast, and thus very low diffraction efficiency. Thus, bio-molecular interactions can be analyzed and quantified by monitoring the intensity of the diffracted light. The major benefit of diffractometric sensors arises from the fact that the fluctuations that limit refractometric sensors (temperature variations, changes in sample composition and non-specific binding) are not spatially coherent, that is, they do not occur in a regular pattern. Thus they contribute very little to the diffracted signal (up to incoherent scattering in all directions, which is unavoidable). Despite this inherent advantage, the field of diffraction based biosensing is not equally well explored as refractometric sensing.

Almost three decades ago, a year after the commercial release of the first surface plasmon resonance device \cite{Liedberg1995-hg}, Tsay \textit{et al.} \cite{Tsay1991-mc} introduced the concept of diffractometric biosensing by detecting choriogonadotropin, a hormonal biomarker, in serum. Further development of the technology enabled the detection of volatile compounds \cite{Bailey2002-sh} and multiplexed signal readout \cite{Goh2002-nr,Goh2005-iq}. Labeled approaches for signal enhancement were developed using gold nanoparticles or enzyme amplification \cite{Goh2003-ut, Loo2005-yu}. These efforts resulted in DotLab$^{\rm{TM}}$, a commercial device based on diffractometric biosensing \cite{Borisenko2006-uv, Gnanaprakasa2011-mp, Pak2014-ki}. Further approaches to enhance the signal were applied by using different optical configurations \cite{Liscidini2007-pk, Lai2008-bs,Yu2004-hr,Yu2004-tw}\footnote{An error in Eq. (2) should be noted: In this equation, $\Lambda$ does not refer to the grating period but to the wavelength of the n=-1 wave inside the grating region in the direction normal to the waveguide surface as described in Ref. \cite{Tamir1977-af}} or by adjusting the diffractive pattern \cite{Jeong2012-ja}. Others have focused on an indirect diffractometric approach, by forming a coherent pattern of a hydrogel, which changed its shape upon target exposure. The shape change resulted in a change of diffraction efficiency \cite{Ye2010-ym,Wang2015-bx, Wang2013-lw}. More recently, the technology has been applied to monitor bacterial growth \cite{Cynthia_Goh2017-wj} or detect low molecular weight organic compounds by means of a competitive assay \cite{Avella-Oliver2018-lh}. In addition, a low-cost diffractometric readout and patterning system based on compact disk technology \cite{Avella-Oliver2017-ju} has been demonstrated. The recently introduced diffractometric biosensing method, focal molography \cite{Fattinger2014-pi,Gatterdam2017-bk, Frutiger2019-um, Frutiger2018-uh},  allows ultra sensitive detection due to focusing of the sensor signal and a photolithographic method for synthesis of a sub-micron diffractive structure of receptors. Due to the inherent robustness of the sub-micron referencing, even interactions in the membrane of living cells can be monitored in real-time with focal molography. \cite{Reichmuth2020-dm}

In spite of its long history, the diffractometric biosensing field did not mature to its full potential and only found limited recognition in the biosensing community. \cite{Gaudin2017-gb}. The reason for this stagnation might be the lack of careful quantitative analysis. Experimental signals are reported in arbitrary units \cite{Avella-Oliver2018-lh}, percental change of intensity \cite{Goh2005-iq}, relative change in diffraction efficiency \cite{Ye2010-ym} or signal to background ratio \cite{Avella-Oliver2018-lh}. This renders the comparison between different arrangements inherently difficult. Moreover, the limit of detection is only measured in values of concentration \cite{Avella-Oliver2018-lh}. Since the response of the sensor to concentration is strongly dependent on the affinity of the recognition element to the target molecule, a comparison between different experiments is hindered. In contrast, the refractometric sensing community uses assay independent units such as surface mass density $\Gamma$ (pg/mm$^2$) and refractive index units (RIU). \cite{Schasfoort2017-hs} We have addressed this issue in our former work on focal molography, by displaying the results as a quantitative mass density modulation $\Delta_\Gamma$ \cite{Gatterdam2017-bk,Fattinger2014-pi}. More recently, we have introduced the more generally applicable coherent surface mass density $\Gamma_{\rm{coh}}$  \cite{Frutiger2019-um, Frutiger2018-uh}. In these publications, we verified that our analytical derivations are in agreement with experimental data. The formalism presented in Ref. \cite{Frutiger2018-uh, Frutiger2019-um} is, however, specific to focal molography and does not directly apply to other diffractometric sensor arrangements. 

In this publication, we extend the results of Ref. \cite{Frutiger2019-um} and present a unifying approach to quantify the signal of any diffractometric biosensor by means of the measured diffraction efficiency and express it via the coherent surface mass density $\Gamma_{\rm{coh}}$. We also describe how to translate $\Gamma_{\rm{coh}}$ into total surface mass density $\Gamma_{\rm{tot}}$ which is an important parameter for many assays \cite{Frutiger2018-uh}. We first explain the necessary theoretical tools and derive the equations describing the diffraction efficiency for a certain amount of adsorbed biological mass. Then, we show how to apply the theory to various configurations of diffractometric sensors, comparing the diffraction from or to guided modes with diffraction between free space modes. This will enable the reader to use the applicable expressions with little extra effort. We hope that this leads to more comparability and accelerates the advancement of the field.
 
\section*{Methods}
\begin{figure}
    \centering
    \includegraphics{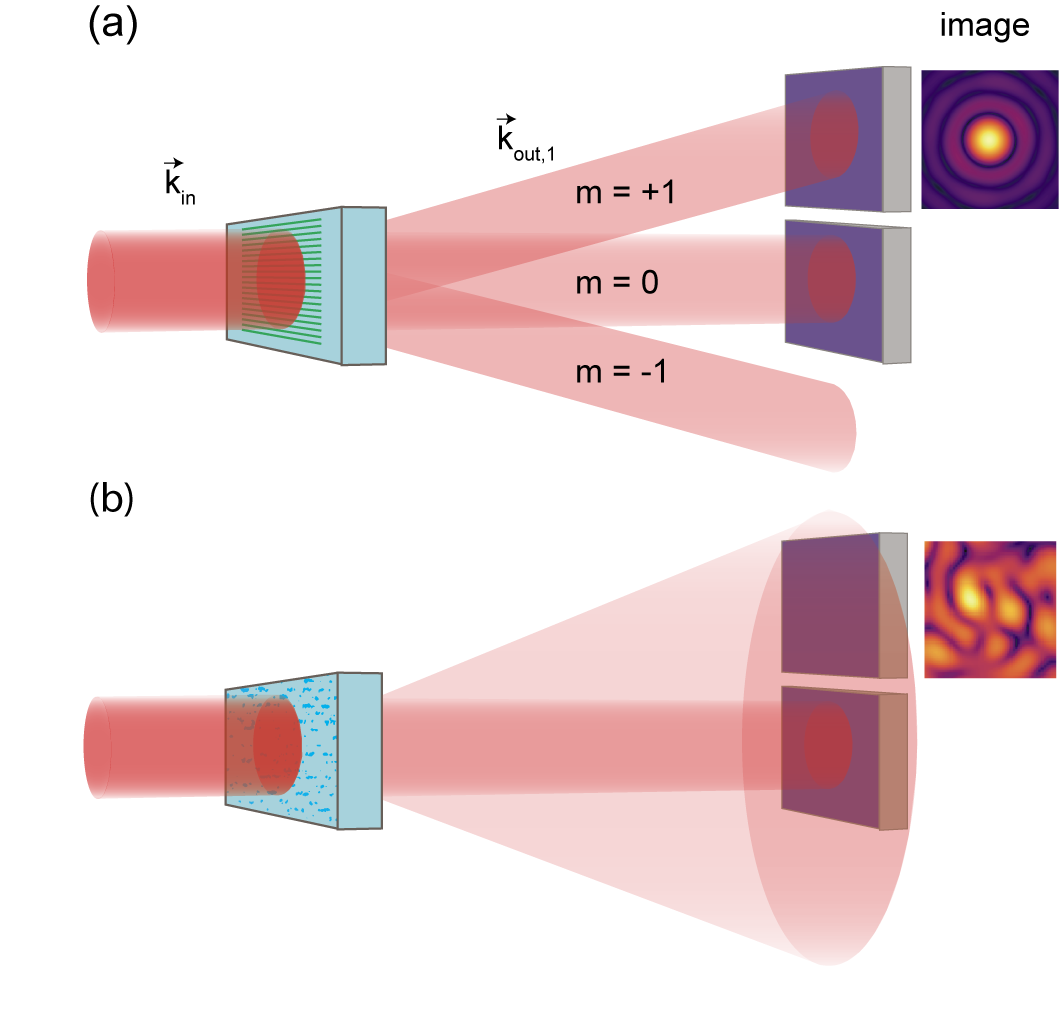}
    \caption{Signal and background in diffractometric biosensors (a) Light is impinging on a molecular grating, which causes diffraction into several orders. The power that is diffracted can be measured by detecting the diffracted light with a detector (camera). A possible image of the diffracted beam is illustrated on the right. It is useful to also monitor the zeroth order to have a power reference, which is required to compute the absolute diffraction efficiency. (b) Non-coherent refractive index distortions such as surface roughness can lead to background light on the camera. Due to the coherent light, this leads to a speckled background in the image and can limit the performance of the sensor.}
    \label{fig:diffractometric sensor}
\end{figure}

Diffractometric sensors are employed in several different optical configurations. A typical setup is illustrated in Figure \ref{fig:diffractometric sensor}a. Light  impinges on the sensor from a certain direction, described by the wavevector $\vec k_{\rm{in}}$. The coherent structure of target molecules diffracts the light into several diffraction orders $\vec k_{\rm{out,}m}$. The electromagnetic power in one of these orders (for maximum resolution the order $m=-1$ or $m=+1$) is measured with a detector and the diffraction efficiency can be deduced. In the following, we present a theory to quantify the adsorbed biological mass density from the diffraction efficiency. In addition to the diffraction from the coherent pattern, non-coherent scattering from random patterns can occur. As mentioned in the introduction, such non-coherent effects are effectively suppressed in diffractometric sensors. Nevertheless, it is important to consider these effects as they lead to a small amount of light being scattered to the detector (Figure \ref{fig:diffractometric sensor}b). This will limit the sensor performance, depending on the detection scheme \cite{Frutiger2019-um}. Thus, we derive equations to quantify both the diffraction from coherent molecular gratings as well as the scattering from random patterns using the coupled mode theory approach based on the ideal mode expansion \cite{Marcuse1974-uo}. This approach has been verified experimentally for one specific sensor configuration in Ref. \cite{Frutiger2019-um}. This paper generalizes this result to all possible two dimensional diffractometric biosensing arrangements. In addition, we provide the analytical expressions to compute the quantification formulas for all arrangements at the interface of two dielectrics.

\subsection*{Coherent surface mass density - The quantitative parameter for diffractometric biosensors}
\label{sec:coherent mass}
Diffractometric biosensors lack a universal parameter for quantification. This prevents the determination of the limit of resolution and hinders the comparison of diffractometric sensors amongst each other or to other sensing techniques quantitatively. A reasonable way to display the sensor output is the diffraction efficiency, which is straightforward to measure. However, the diffraction efficiency depends on the geometry and the optical configuration of the sensor which again hinders comparison of the actual sensor performance. Instead, an expression for the actual quantity of interest, the adsorbed mass is required.

In contrast to the diffractometric biosensing field, the use of such a quantity is common in refractometric biosensors and surface plasmon resonance in particular. For quantification, results are displayed either as refractive index units or as surface mass density, the former being more general. The exact conversion between these two depends on the configuration, but a rule of thumb is $10^{-6} \text{ RIU} \approx 1 \text{ pg/mm}^2$  for proteins and surface plasmon resonance \cite{Homola2008-px, Piliarik2009-xb}.

For diffractometric sensors it is not straightforward to define a surface mass density as the biomolecules adsorb in a periodically modulated manner across the sensor. A possible solution is to use the surface mass density modulation $\Delta_\Gamma = \Gamma_+ - \Gamma_-$ as we did in earlier publications \cite{Gatterdam2017-bk, Fattinger2014-pi}. This value measures the difference between the surface mass density of active and passive regions. Unfortunately, this parameter is not universally applicable since the signal also depends on the mass distribution. Therefore, the same $\Delta_\Gamma$ can result in different signals depending on the exact mass distribution. In addition, the definition of active and passive regions is ambiguous which can cause complications.

\begin{figure}
    \centering
    \includegraphics{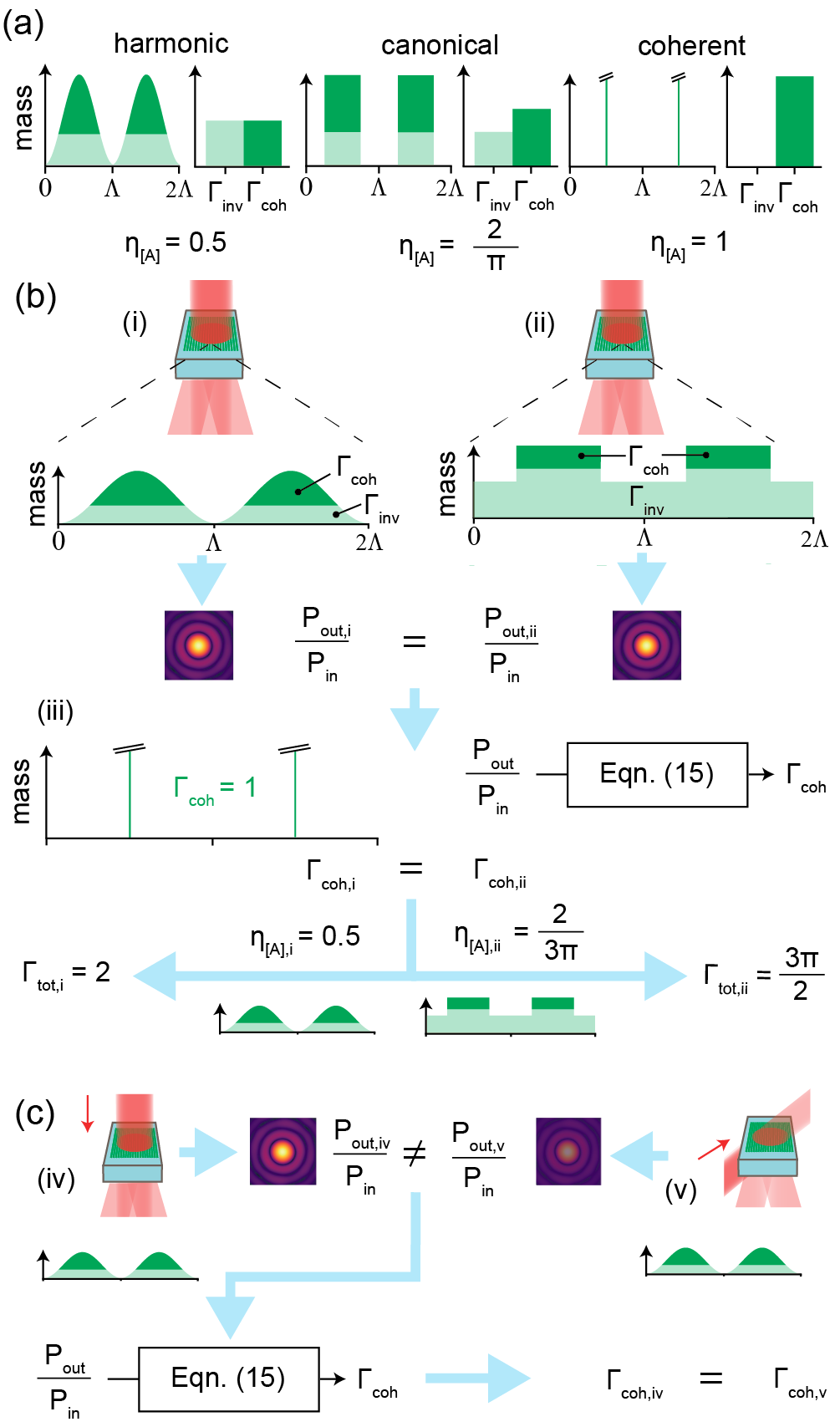}
    \caption{(a) Illustration of the same total mass density $\Gamma_{\rm{tot}}$ in three different distributions (harmonic, canonical and coherent). The different distributions vary in their proportion of $\Gamma_{\rm{coh}}$ and $\Gamma_{\rm{inv}}$. Therefore, all three distributions cause a different signal. (b) Illustration of how to use the theory described in this paper. The two sensors (i) and (ii) have a different total amount of immobilized mass and the mass is distributed differently across the sensor area. Nevertheless they both have the same diffraction efficiency $ \frac {P_{\rm{out}}}{P_{\rm{in}}}$. In order to address this properly the coherent surface mass density, $\Gamma_{\rm{coh}}$, is used. $\Gamma_{\rm{coh}}$ is defined as the amount of mass that causes the measured $\frac{P_{\rm{out}}}{P_{\rm{in}}}$ if the mass is arranged as illustrated in (iii). $\Gamma_{\rm{coh}}$ can be calculated from Eqn. \eqref{eq:universal_mass_quantification}. It is equal for sensor (i) and (ii) as both have the same optical configuration and the same $\frac{P_{\rm{out}}}{P_{\rm{in}}}$. When the mass distribution is known (for example by investigation of the manufacturing process) the analyte efficiency $\eta_{[A]}$ can be determined. It can be used to calculate the total mass density from $\Gamma_{\rm{coh}}$. (c) illustrates that $\Gamma_{\rm{coh}}$ can be determined independently of the optical set-up. The two sensors (iv) and (v) have the same amount and distribution of adsorbed mass, but use a different optical configuration. This results in a different $\frac{P_{\rm{out}}}{P_{\rm{in}}}$, but $\Gamma_{\rm{coh}}$ determined from Eqn. \eqref{eq:universal_mass_quantification}  remains the same. This allows us to compare the sensor output from different optical configurations.}
    \label{fig:surface mass modulationl}  
\end{figure}

A generally applicable parameter for diffractometric biosensors is the \textit{coherent surface mass density}, $\Gamma _{\rm{coh}}$. $\Gamma _{\rm{coh}}$ is the surface mass density required to cause the measured diffraction efficiency under the assumption that the mass is arranged perfectly coherently (\textit{i.e.} at the center of the constructively interfering regions) (Figure \ref{fig:surface mass modulationl}). Mathematically, it is the Fourier coefficient that corresponds to the grating period of the mass distribution function $\Gamma\left(x,y\right)$ 
\begin{equation}\label{eq:gamma_coh}
{\Gamma _{{\rm{coh}}}} = {{\int {\int\limits_A {\sin \left( {{{\vec \beta }_g} \cdot \vec r} \right)\Gamma \left( {x,y} \right)dxdy} } } \over A},
\end{equation}
where the grating vector $\vec \beta_g$ fulfills the resonance condition for diffraction. We choose the vector to be along the $x$-axis such that $\vec \beta_g=\left(\frac{2\pi}{\Lambda},0,0\right)$, where $\Lambda$ is the grating period and $\vec{r}\equiv(x,y,0)$). $\Gamma _{\rm{coh}}$ can be calculated directly from the diffraction efficiency. It is independent of the mass distribution and can be used to compare different sensors or assays. Thus, the coherent surface mass density is what diffractometric biosensor can "see". This implies that there is a surface mass density that diffractometric biosensors cannot measure,  \textit{i.e.} the \textit{invisible surface mass density} $\Gamma_{\rm{inv}}$. The total mass density of molecules that is bound to the immobilized receptors on the sensor surface,

\begin{equation}\label{eq:gamma_tot}
    {\Gamma _{{\rm{tot}}}} = {{\int  \int\limits_A {\Gamma \left( {x,y} \right)dxdy} } \over A},
\end{equation}
can be split into these two distinct components: the invisible surface mass and the coherent surface mass such that $\Gamma_{\rm{tot}}=\Gamma_{\rm{coh}} + \Gamma_{\rm{inv}}$. Figure \ref{fig:surface mass modulationl}a illustrates the two components schematically for three prototypic diffractometric mass modulations, the harmonic modulation, the rectangular (or canonic) modulation and the fully coherent modulation of the sensing structure. The invisible surface mass consists of two contributions: A homogeneous background of evenly distributed receptors and a second contribution arising from the non-coherent fraction of the modulated distribution of receptors on the sensor surface. The ratio of the invisible surface mass and the coherent surface mass is determined by the (photolithographic) synthesis process of the sensing structure. Ideal synthesis processes create sensing structures with only a small homogeneous background surface mass density compared to the modulated part of the distribution of immobilized receptors on the sensor surface. If the mass distribution within a unit of the spatially periodic arrangement is known, the measured coherent surface mass density can be converted seamlessly to the total surface mass density \cite{Frutiger2019-um, Frutiger2018-uh}
\begin{equation}
    \label{eq:analyte efficiency}
    \Gamma_{\rm{tot}} = \frac{\Gamma_{\rm{coh}}}{\eta_{[A]}} .
\end{equation}
$\Gamma_{\rm{tot}}$ is an important parameter for many applications, since it allows to compute receptor occupancies \cite{Frutiger2018-uh}. The analyte efficiency $\eta_{[A]}$ is a characteristic value of the mass distribution (which could be determined, e.g., by detailed analysis of the manufacturing process). Its minimum value is 0 and is maximum value is 1, whereas currently achieved experimental values reach about 0.25. \cite{Frutiger2018-uh} The analyte efficiency $\eta_{[A]}$ describes how efficiently the analyte molecules diffract the light if arranged in a certain distribution (Figure \ref{fig:surface mass modulationl}b) and can be viewed as an analogue of the structure factor in crystallography. The analyte efficiency can be computed for a certain mass distribution $\Gamma(x,y)$ by \cite{Frutiger2019-um}
 \begin{equation}\label{eq:analyte_efficiency_analytical}
    {\eta _{\left[ {\rm{A}} \right]}} = \frac{\Gamma_{\rm coh}}{\Gamma_{\rm tot}}={{\int  \int\limits_A {\sin \left( {{{\vec \beta }_g} \cdot \vec r} \right)\Gamma\left( {x,y} \right)dxdy} } \over {\int  \int\limits_A {\Gamma\left( {x,y} \right)dxdy}}}.
 \end{equation}
  Values of $\eta _{\left[ {\text{A}}\right]}$ for some example distributions are given in Ref. \cite{Frutiger2019-um}. Higher diffraction orders will have a lower analyte efficiency and a higher fraction of invisible surface mass density. Therefore, one should always measure the first diffraction order. The simple conversion from $\Gamma_{\rm{coh}}$ to $\Gamma_{\rm{tot}}$ allows straightforward comparisons to other label-free sensors. 

\subsection*{Coupled Mode Theory}
The universal diffractometric sensor parameter $\Gamma_{\rm{coh}}$ is directly linked to the diffraction efficiency in any sensor configuration. Although it is straightforward to measure the diffraction efficiency experimentally, it is challenging to connect it mathematically to $\Gamma_{\rm{coh}}$. In the following, we explain how to determine it using coupled mode theory.

\begin{figure}
    \centering
    \includegraphics{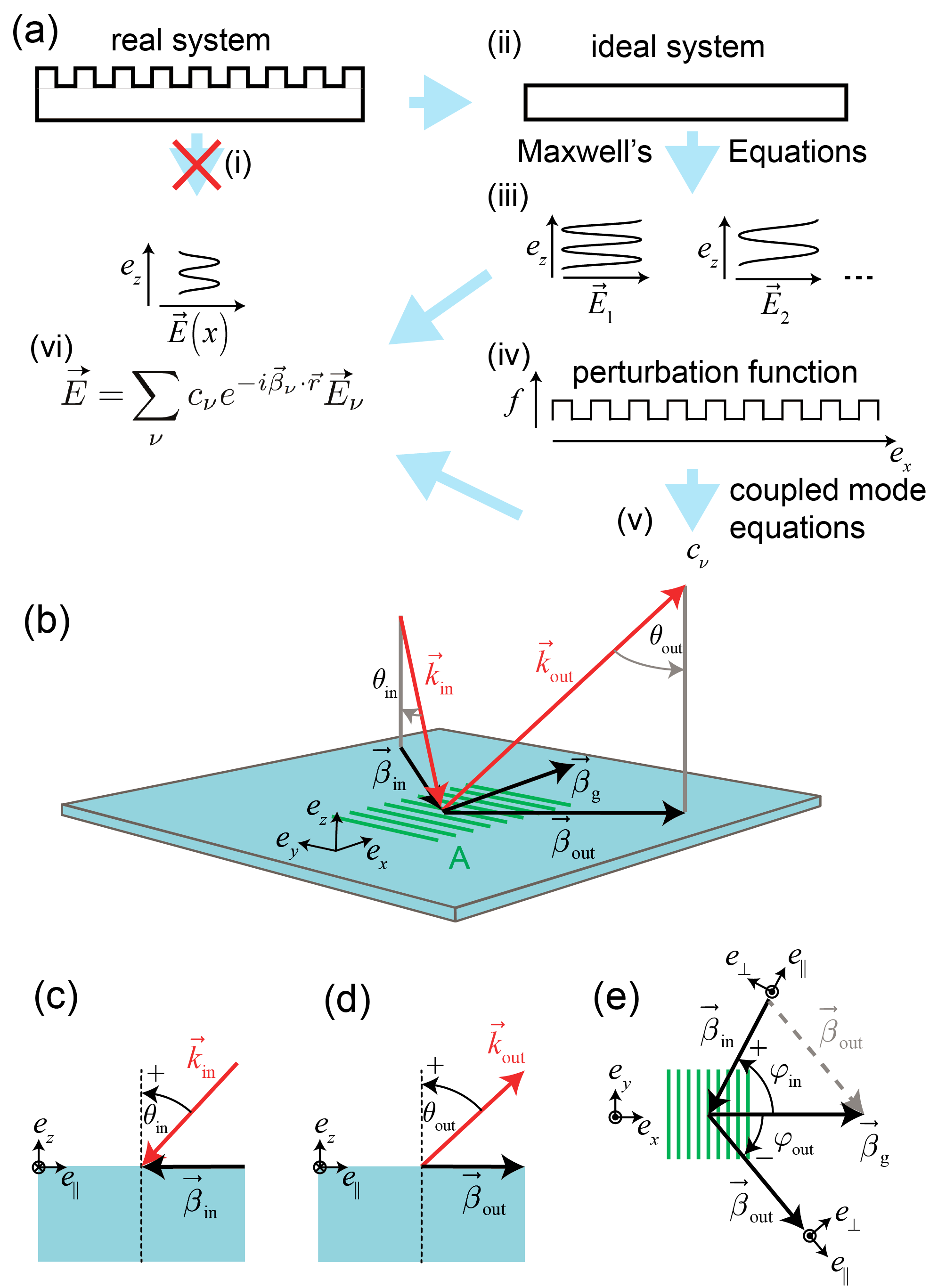}
    \caption{(a) Illustration of the ideal mode expansion. (i) The electromagnetic field of a real system is often difficult to solve for by applying Maxwell's equations directly. Therefore, ideal mode expansion is applied. (ii)  A simpler ideal system is taken as reference, for which Maxwell's equations can be solved easily to yield a complete set of modes ((iii)  $\vec{E}_{1}$, $\vec{E}_{2}$ ...) (iv) The difference between the ideal and the real system can be described by a perturbation function in a spatially narrow region close to the interface at $z=0$. This perturbation leads to energy transfer between the ideal modes of the system - \textit{i.e.} coupling (v) By solving the coupled mode equations the expansion coefficient for each ideal mode $c_\nu$ can be calculated. (vi) From the ideal modes and the corresponding expansion coefficients $c_\nu$ the real electromagnetic fields can be determined using a linear combination. (b) General representation of the notation and coordinate system used throughout this paper. The vectors $\vec \beta_{\rm{in}}$ and $\vec \beta_{\rm{out}}$ are projections of the k-vectors of the incident $\vec k_{\rm{in}}$ and the outgoing mode $\vec k_{\rm{out}}$ onto the $xy$-plane defined by the boundary of two dielectrics. The $x$-axis is chosen normal to the grating lines (parallel to the grating vector $\vec \beta_{\rm{g}}$). The molecular grating has an area $A$. The incident and outgoing k-vector together with the surface normal define the plane of incidence and the outgoing plane. (c) plane of incidence: The incident k-vector is inclined with respect to the surface normal by an angle $\theta_{\rm{in}}$. The angle $\theta$ is always measured from the k-vector to the surface normal and is positive in counter clockwise direction when looking in positive $e_{\bot}$-direction. (d) outgoing plane: The outgoing k-vector is forms an angle $\theta_{\rm{out}}$ with the surface normal. (e) Top view onto the coupling plane ($xy$) illustrating the coupling condition: Plane of incidence and outgoing plane are rotated with an azimuth $\varphi_{\rm{in}}$, $\varphi_{\rm{out}}$ with respect to the positive grating normal direction. $\varphi$ is again positive in counter clockwise direction when looking in negative $z$-direction.}
    \label{fig:ideal mode expansion}
\end{figure}

Coupled Mode theory is a perturbation theory that is used to describe different optical elements such as grating couplers \cite{Norton1997-nl} or directional couplers \cite{Huang1994-hr}. It is based on the fact that an electromagnetic field configuration can be written as a linear combination of orthogonal basis functions (modes). Modes are linearly independent monochromatic (single-frequency) solutions of Maxwell's equations in a particular system. A first step in any coupled mode theory implementation is to define a complete orthogonal set of modes. Ideal mode expansion is one possible approach to do so, which is particularly well suited to the optical geometry in diffractometric sensors and results in relatively simple analytical expressions \cite{Marcuse1974-uo}. Its principle is illustrated in Figure \ref{fig:ideal mode expansion}a with the example of a planar dielectric substrate with a diffraction grating etched into its surface. For the full system, it is complicated to solve Maxwell's equations and to determine the electromagnetic fields. Therefore, one considers the real system as a perturbation of an ideal system, in this case, a planar interface (Figure \ref{fig:ideal mode expansion}a), which allows for Maxwell's equations to be solved more easily. In our case, by virtue of the translational invariance of the ideal system in the $xy$-plane, the solutions of Maxwell's equations can be labeled by in-plane momentum $\vec{\beta}$ and polarization and propagation direction perpendicular to the plane for modes propagating freely in all three dimensions. Those solutions provide a complete orthogonal set of modes, into which the electromagnetic field of the real system can be expanded \cite{Marcuse1974-uo}. Mathematically,

\begin{equation}
    \label{eq:ideal mode expansion}
    \begin{split}
    \vec{E} =& \sum_{\nu} c_{\nu} e^{-i \vec \beta_{\nu} \cdot \vec r} \vec{E}_{\nu}(z) \\ 
    \vec{H} =& \sum_{\nu} c_{\nu} e^{-i  \vec \beta_{\nu} \cdot \vec  r} \vec{H}_{\nu}(z),
    \end{split}
\end{equation}
where $c_{\nu}$ is the expansion coefficient of mode $ \nu $. \footnote{$c_{\nu}$ varies slowly in space (on length scales much larger than $1/|\vec{\beta}_\nu|$), $ c_{\nu} = c_{\nu}(x,y)$. Note that the set of modes is labelled by a continuum of in-plane wave-vectors $\vec{\beta}_\nu$, and thus the sums must actually be thought of as integrals.} It should be noted that compared to Ref. \cite{Marcuse1974-uo}, Eqn. (\ref{eq:ideal mode expansion}) has been generalized to two-dimensions transverse to the $z$-axis (the normal to the interface), with $ \vec \beta_{\nu} $ being the projection of the wavevector $ \vec k $ onto the $xy$-plane, $\vec r \equiv (x,y)$ the position vector in that plane (see Figure \ref{fig:ideal mode expansion}b). The definitions used in this paper for the plane of incidence, the outgoing plane, as well as the sign conventions for the angle of incidence $\theta_{\text{in}}$, the outgoing angle $\theta_{\text{out}}$ and the azimuth $\varphi$ are defined in Figure \ref{fig:ideal mode expansion}c,d,e. Upon proper normalization of the modes, the expansion coefficient $c_{\nu}$ is directly related to the power carried by the mode (Ref. \cite{Marcuse1974-uo} Eqn. (5.2-1)).
\begin{equation}
\label{eq:Power}
     P_{\nu}  = {\left| {c_{\nu}} \right|}^2,
\end{equation}
where $||$ indicates the absolute value. 

The  difference between the real and the ideal system is treated as a perturbation. In an ideal system, due to their mutual orthogonality the modes are uncoupled and thus cannot exchange energy between one another. The perturbation, however, induces a coupling between the modes and leads to an exchange of energy. This is described by coupled mode theory. In order to determine the diffraction efficiency of coupling between certain modes one can simply apply Eqn. (\ref{eq:Power}) and analyze the (spatially averaged) ratio between the expansion coefficients of the incident and the diffracted modes. These expansion coefficients can be obtained from solving the coupled mode equations.

The coupled mode equations can be derived from the expansion given in Eqn. (\ref{eq:ideal mode expansion}), the orthogonality relation and Maxwell's equations \cite{Marcuse1974-uo}. In general, they are relatively involved differential equations. However, for most diffractometric sensors several assumptions apply that significantly simplify the analysis. Essentially, the expansion coefficients turn out to be given by a suitable two dimensional Fourier transform over the diffraction area where the perturbation is located (Supplementary Information Section \ref{sec:examples_of_CMT_for_two_dimensions}). We assume that the incident beam is much wider than the molecular grating and furthermore that only two modes need to be considered, an incident beam $c_{\rm{in}}$ and a diffracted beam $c_{\rm{out}}$. \footnote{We change the nomenclature from mode to beam. By a beam we refer to a wavepacket of modes of finite extent, that is, a mode multiplied with a transverse envelope function (see Supplementary Information Section \ref{sec:normalizing_power_for_coupling_coefficient_calculation}), whereby we neglect that Maxwell's equations imply that the envelope function slightly evolves under the propagation of the beam.} Before coupling, there is only the incident beam. The coupling between the two beams is weak such that $c_{\rm{in}}$ can be considered constant. As mentioned above, the coupling between modes is caused by a perturbation of the ideal waveguide. In a diffractometric biosensor, this perturbation is represented by a molecular diffraction grating. Additionally, the effect of random perturbations such as a surface roughness should be considered since these effects might limit the sensor performance.

For a molecular grating as well as for surface roughness we model the perturbation with a constant refractive index, but with a varying height. The perturbation is located on a dielectric interface (at $z=0$) and extends by a height ${f(x,y)}$ into the cover ($z>0$).  ${f(x,y)}$ is called the perturbation function and has an unlimited extent in the $x,y$ plane (Figure \ref{fig:ideal mode expansion}a). Nevertheless, we confine the perturbation to a certain area of interest $A$ which is usually the sensor area. We further assume that ${f(x,y)}$ is much smaller than the wavelength, so that we can consider the electric field to be constant over the entire perturbation in the direction normal to the surface. With these assumptions and using Eqn. \eqref{eq:Power} one can derive that the resulting diffraction efficiency is proportional to the two-dimensional Fourier transform of the perturbation function,
\begin{equation}
\label{eq:general diffraction efficiency integral}
\begin{split}
{{{P_{\rm{out}} }} \over {{P_{{\rm{in}}} }}} =&  {{{\left| {{{{c_{\rm{out}} }} \over {{c_{{\rm{in}}} }}}} \right|}^2}} \\
 = &\left\langle {{{\left| {{{K} } \left( {{n_m^2} - n_c^2} \right) \int\!\!\!\int\limits_A {f(x,y){e^{ i\left( {{{\vec \beta }_{{\rm{in}}} } -  {{\vec \beta }_{\rm{out}} }} \right)\vec r}}dA} } \right|}^2}} \right\rangle.
\end{split}
\end{equation}
Here $\langle\rangle$ indicates an ensemble average. This only becomes important when $f$ is a random function such as the surface roughness, but can be ignored otherwise. The refractive index within the perturbed region is $n_c$ for the ideal configuration, but $n_m$ for the real configuration. The parameter $K$ %$\left|K\right|$
is the coupling coefficient. Only its absolute value $\left|K\right|$ is of physical interest. In our choice of convention $\left| K \right|$ only depends on the optical configuration, and on the cross-section of the incident beam, 
\begin{equation}
\label{eq:constant coupling coef}
\begin{split}
\left|K\right| = & \left| {{ \frac{\omega {\varepsilon _0}} {4P}{}}} \times \right.\\
& \left.\left( {\vec {E}_{\text{out,t}}^{*} (0) \cdot \vec {E}_{\text{in,t}}  (0) + {\frac {n_c^2} {{n_m^2}}}\vec {E}_{\text{out,n}}^{*} (0) \cdot \vec {E}_{\text{in,n}} (0)} \right)\right|.
\end{split}
\end{equation}
Here, $\omega$ is the angular frequency of the mode, $\varepsilon_0$ is the permittivity of vacuum and $\vec {E}_{\text{in,t}}$, $\vec {E}_{\text{in,n}}$ are the electric field components tangential and normal to the dielectric interface of the normalized incident mode, and analogously for the outgoing mode $\vec {E}_{\rm{out}}$. 
$P$ is the power a normalized beam has when its expansion coefficient is set to $c=1$. The incident and outgoing power are thus given by ${P_{{\rm{out}}}} = \left|{c_{{\rm{out}}}}\right|^2P$ and ${P_{{\rm{in}}}} = \left|{c_{{\rm{in}}}}\right|^2P$. Since the fields $\vec{E}_{\rm in, out}\sim \sqrt{P}$ scale as the square root of the power, the arbitrary choice of $P$ drops out form Eq.~(\ref{eq:constant coupling coef}).\footnote{In \cite{Fattinger2014-pi, Marcuse1974-uo} modes were infinitely extended in the third dimension perpendicular to the propagation direction. Thus they were normalized in terms of power per unit cross-section in the third direction. Here instead we treat beams of finite width which are thus normalized to have a finite power.}
The electric fields are evaluated at $z =0^+$ right on top of the interface, on the side of the cover. The coupling coefficient $\left|K\right|$ essentially captures the overlap of the fields of the incident and the outgoing beams at the dielectric interface and thereby determines the coupling strength between two modes. The field strengths are obtained by solving Maxwell's equations for the ideal optical system of the diffractometric sensor and normalizing the fields (see Supplementary Information). \footnote{It should be noted that the definition of $\left|K\right|$ in this paper is different from the one used in Eqn. 3.4-7 of Ref. \cite{Marcuse1974-uo}. In addition, small adjustments with respect to Ref. \cite{Marcuse1974-uo} have been made to correct for shortcomings of the ideal mode expansion approach in the case of transverse magnetic (TM or p) polarization (see Supplementary Information Section \ref{sec:two_dimensional_coupled_mode_theory}) \cite{Hall1991-tu, De_Sterke1990-lz}.}

%We will do so for a sinusoidal perturbation, but the result will be generally applicable to any perturbation thanks to the analyte efficiency $\eta_{[A]}$.

\subsection*{Application of coupled mode theory to periodic and random perturbations} 

So far, we have explained the principle of coupled mode theory and given the fundamental formulas for describing diffractometric sensors. We will now illustrate how these expressions can be applied to analyze periodic and random perturbations. We first show how to treat periodic molecular diffraction gratings. Then, we will address a random perturbation, such as surface roughness, which may be limiting the performance of the sensor \cite{Frutiger2019-um}. 

\begin{figure*}
    \centering
    \includegraphics{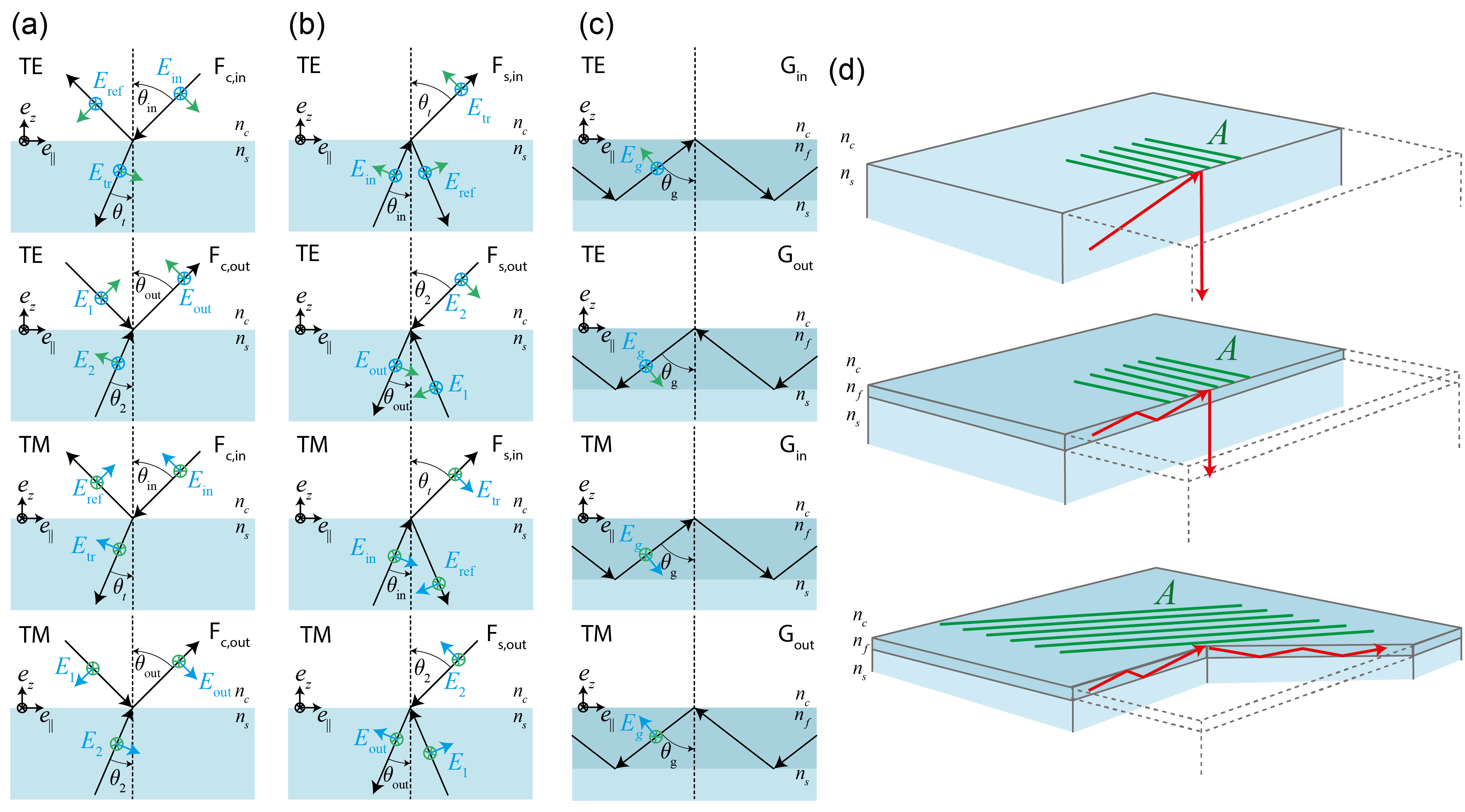}
    \caption{Modes of the ideal interface structure used in the description of two-dimensional diffractometric biosensors. Freely propagating (F  in (a,b)) and guided modes (G in (c)) are distinguished. The F modes are further distinguished according to whether they originate or end up in the cover  (F$_{c}$ (a)) or in the substrate (F$_{s}$ (b)), respectively. Every mode is either incident or outgoing, and has a TE (s) or a TM (p) polarization. 
    The direction of the electric (blue) and magnetic (green) fields is illustrated for the chosen convention, which results in the signs appearing in Table \ref{tab:normalized_field_components}. For a particular mode, the field intensities in the different media are related by the Fresnel coefficients following the definition in \cite{Novotny2011-pa}. The fields of the outgoing modes can be found by time reversal of the in-arrangement (see Supplementary Information \ref{Example normalization of Fcout for TE polarization}). The angles of the field vector for the guided mode are connected to the effective refractive index of the mode $N = {n_f}\sin \left( {{\theta _{\rm{g}}}} \right)$. The fields involved in the coupling at the molecular grating are evaluated in the cover and their orientation needs to be determined from Snell's law and its generalization in the case of total internal reflection for beams incident from below the cover (F$_s$ and G). \cite{Novotny2011-pa} (d) Three exemplary optical configurations for diffractometric sensors on a planar substrate. They couple different incident and outgoing modes (indicated by red arrows): Top: Coupling between two F$_s$ modes. Middle: Coupling between a G and an F$_s$ mode. Bottom: Coupling between two G modes with different propagation directions in the waveguide plane.}
    \label{fig:optical configurations}
\end{figure*}

\subsubsection*{Periodic perturbation functions}

In diffractometric biosensing, the signal is caused by a periodic perturbation function $f(x,y)$ of constant refractive index $n_m$. To arrive at a general description of the diffraction efficiency as a function of the bound mass, $n_m$ and $f\left(x,y\right)$ must be linked to the mass density distribution $\Gamma\left(x,y\right)$. This link is derived in Supplementary Section \ref{sec:Conversion of refractive index to coherent mass} in Eqn. \eqref{eq:link_between_mass_density_and_perturbation}. Eqn. \eqref{eq:general diffraction efficiency integral} can then be written as a function of the mass density distribution $\Gamma\left(x,y\right)$

\begin{equation}\label{eq:periodic_perturbation_mass}
    {{{P_{{\rm{out}}}}} \over {{P_{{\rm{in}}}}}} = {\left| {2{n_c}{{dn} \over {dc}}K\int  \int\limits_A {\Gamma \left( {x,y} \right){e^{i\left( {{{\vec \beta }_{{\rm{in}}}} - {{\vec \beta }_{{\rm{out}}}}} \right)\vec r}}dA} } \right|^2}.
\end{equation}
The refractive index increment ${dn} \over {dc}$ describes the variation in refractive index with the solute concentration. For proteins in water $\frac{d n}{d c}= 0.182\, {\rm ml/g}$ is a commonly accepted value \cite{De_Feijter1978-bx}. To evaluate the integral we write the mass density distribution as a Fourier series, 
\begin{equation}
\label{eq:fourier series}
    \Gamma \left( {x,y} \right) = \sum\limits_j {{a_j}{e^{i\left( {{{\vec \beta }_j} \cdot \vec r} \right)}}},
\end{equation}
and consider the resonance condition, \linebreak $\left( {{{\vec \beta }_{{\rm{in}}} } -  {{\vec \beta }_{\rm{out}} }} \right) \to \left(\frac{2\pi }{\Lambda},0,0\right)=\vec{\beta}_g$ when diffraction is most efficient. Upon substitution of Eqn. \eqref{eq:fourier series} into Eqn. \eqref{eq:periodic_perturbation_mass} the integrand $\Gamma \left( {x,y} \right){e^{i\vec\beta_g\vec r}} = {a_g} + \sum\limits_{j \ne g} {{a_j}{e^{i\left( {\left( {{{\vec \beta }_j} + {{\vec \beta }_{\rm{g}}}} \right) \cdot \vec r} \right)}}} $ consists mostly of terms that are oscillating quickly over the integration area. The integral of these terms is therefore negligible. Only the term $\Gamma_{\rm{coh}}:= a_g$ remains, which corresponds to a spatial frequency of $\vec \beta_i = -\vec \beta_g$. This holds as long as the integration area is significantly larger than the grating period. At the resonance condition the integral therefore evaluates to
\begin{equation}
\int  \int\limits_A {\Gamma \left( {x,y} \right){e^{i\beta_g\vec r}}dA} = {\Gamma _{{\rm{coh}}}}A
\end{equation}
and Eqn. \eqref{eq:periodic_perturbation_mass} can be written as
\begin{equation}\label{eq:periodic_perturbation_evaluated}
{{{P_{{\rm{out}}}}} \over {{P_{{\rm{in}}}}}} = 4{\left|K\right|}^2n_c^2{\left( {{{dn} \over {dc}}} \right)^2}\Gamma _{{\rm{coh}}}^2  A^2.
\end{equation}
This equation connects the diffraction efficiency and the coherent surface mass density for any diffractometric sensor and general periodic mass distribution functions $f$. The only parameter that needs to be computed for a particular arrangement is the coupling coefficient $\left|K\right|$. Eqn. \eqref{eq:periodic_perturbation_evaluated} is valid under certain assumptions that are met in most biosensing application.\footnote{We have assumed that the incident beam is wider than the molecular grating and that its intensity profile is uniform across the grating. Furthermore, the coupling is assumed to be weak, a condition which should always be met for biological gratings.} The expression has been derived at the resonance condition. For a more general solution independent of the resonance condition see SI Section \ref{sec:examples_of_CMT_for_two_dimensions}. Finally it should be noted that, in order to estimate the total bound mass, the analyte efficiency $\eta_{[A]}$ needs to be determined experimentally for the particular system \cite{Frutiger2019-um}.

\subsubsection*{Random perturbation function}
Non-coherent effects can lead to significant background scattering as illustrated in Figure \ref{fig:diffractometric sensor}b and can therefore limit the sensor performance \cite{Frutiger2019-um}. An example for such an effect is scattering at surface roughness. Surface roughness can be described by a random perturbation function $f(x,y)$ with an exponentially decaying radially symmetric autocorrelation function $g\left(r\right):=\langle f(r')f(r' + r) \rangle$ 
\begin{equation}
\label{eq:autocorrelation}
g(r) = {\sigma ^2}{e^{ - (r/{L_c})}},
\end{equation}
where $\sigma$ is the root mean square roughness, $L_c$ is the correlation length and $r$ is the radial distance. The root mean square roughness $\sigma$ is determined by the height fluctuations of the dielectric interface. The correlation length $L_c$ characterizes their typical lateral size. In all practical cases, the correlation length $L_c$ is much smaller than the sensor dimensions. It can be shown that for this case, the diffraction efficiency given in Eqn. (\ref{eq:general diffraction efficiency integral}) can be written as (Supplementary Information Section \ref{sec:roughness_perturbation_function})
\begin{equation}
\label{eq:diffraction efficiency roughness}
{{{P_{\rm{out}} }} \over {{P_{{\rm{in}}} }}} = {A{ \left| {K} \right|^ 2}}\left( {{n_d^2} - n_c^2} \right)^2 {{2\pi {\sigma ^2}L_c^2} \over {{{\left( {1 + {|\vec{\beta_g}|^2}L_c^2} \right)}^{3/2}}}},
\end{equation}
where $n_d$ is the refractive index of the dielectric that comprises the roughness. $n_c$ is the refractive index of the cover and $|\vec{\beta_g}|=\frac{2\pi}{\Lambda}$ is the modulus of the grating wavevector (which equals the transferred in-plane momentum).

\section*{Results}
The content of the theoretical results described in the methods is elucidated in the following section. After stating the formula for a generally applicable quantification using the coherent surface mass density, $\Gamma_{\rm{coh}}$, we explain how to calculate the coupling coefficient for different optical configurations which enter that formula. To enable straightforward implementation the necessary expressions for all configurations have been derived and are summarized in a table. Three cases are calculated in detail as examples. We then explain how background scattering should be accounted for to estimate sensor performance. Finally, we end with a discussion about a proper characterisation of diffractometric sensors.

\subsection*{Coherent surface mass density quantification}
We have defined $\Gamma_{\rm{coh}}$ to be the universal parameter to quantify the performance of diffractometric sensors. 
Using Eqn. (\ref{eq:periodic_perturbation_evaluated}) it can be determined by measuring the diffraction efficiency ${{{{P_{{\rm{out}}}}} \over {{P_{{\rm{in}}}}}}}$ for a general optical configuration,
\begin{equation}
{\Gamma _{{\rm{coh}}}} = {1 \over {2\left| K \right|{n_c}{{d n} \over {d c} }A}}\sqrt {{{{P_{{\rm{out}}}}} \over {{P_{{\rm{in}}}}}}}, 
\label{eq:universal_mass_quantification}\end{equation}
where $A$ is the sensor area, $\frac{dn}{dc}$ is the refractive-index increment and $n_c$ is the refractive index of the cover medium. The optical configuration is described by the coupling coefficient $\left|K\right|$ which is computed for specific cases in the next subsection. Eqn. \eqref{eq:universal_mass_quantification} is the universal mass quantification formula for an arbitrary diffractometric biosensor. For completeness, we also stated the universal number density quantification formula for particles with polarizability $\alpha$ in the Supplementary Information Section \ref{sec:universal_number_density_quantification}.d

\subsection*{Coupling coefficients $K$ for different optical configurations}
In order to quantify $\Gamma_{\rm{coh}}$ of a given diffractometric biosensor configuration, the coupling coefficient $\left|K\right|$ between the incident and the outgoing mode needs to be calculated, for which we provide a simple recipe applicable to almost any diffractometric biosensor on a 2D support. For each optical configuration the molecular grating is situated at an interface between a support and a cover medium. We consider configurations where the support consists of a substrate with refractive index $n_s$. Alternatively, the support can consist of a waveguide film with refractive index $n_f$ on a substrate with refractive index $n_s< n_f$ allowing for guided incident or outgoing modes. The cover medium has refractive index $n_c$. 

A coupling configuration is defined by the type of incident and the outgoing modes. 
For freely propagating modes (F) one needs to distinguish whether they originate from the cover ($F_c$) or from the substrate $F_s$) (for incoming modes) or whether they propagate towards the cover or the substrate, respectively, for outgoing modes.
If a waveguide is present, modes can be guided (G) and thus be tied to the interface region.\footnote{As a side remark, at metal dielectric interfaces surface plasmon polaritons can occur as guided modes and can be described with this formalism as well \cite{Yu2004-tw}. However, the expressions for the fields on the surface (Table \ref{tab:normalized_field_components}) have not been computed for surface plasmons. For the field distribution of plasmons see \cite{Lukosz1991-wi}. } For all cases (F$_c$, F$_s$ and G),  two possible polarizations, transverse electric TE (\textit{i.e} s-polarization), and transverse magnetic TM (\textit{i.e.} p-polarization), can be considered.  This results in 6 possibilities for in-and outgoing modes, respectively, (see Figure \ref{fig:optical configurations}a,b). Once the incident and outgoing mode of a diffractometric sensor configuration have been selected their field overlap must be determined to calculate the coupling coefficient $\left|K\right|$,  using Eqn. \eqref{eq:constant coupling coef}. We rewrite it in a different form here to allow for more straightforward implementation
\begin{equation}
\label{eq:constant coupling coef application}
\begin{split}
\left|K\right| =  \left|{{ \frac{\pi  } {2 \lambda Z_0 P}}}\right.  & \left(  \vphantom {\frac {n_c^2} {{n_m^2}}} {E}_{\text{out},x}^{*}(0) {E}_{\text{in},x}(0) + {E}_{\text{out},y}^{*} (0) {E}_{\text{in},y} (0) \right. \\
& \left.\quad \left.{}  + {\frac {n_c^2} {{n_m^2}}}{E}_{\text{out},z}^{*} (0) {E}_{\text{in},z} (0) \right)\right|.
\end{split}
\end{equation}
$Z_0 = \frac{1}{\varepsilon_0 c_0}$ is the impedance of vacuum, $c_0$ the speed of light in vacuum, $\lambda$ the wavelength in vacuum and $P$ is the power used to normalize the beam fields. Both $Z_0$ and $P$ cancel out when the normalized electric field components $E_{\rm{in/out,i}}$ are substituted by the respective expressions given in Table \ref{tab:normalized_field_components}. The derivation of the normalized beam fields is given in the Supporting Information Section \ref{field normalizations}.
The free space modes (F) are defined by the incident or outgoing angles, respectively, the wavelength and the refractive indices of the media. The fields of guided modes on the other hand are characterized by the effective thickness $t_{\rm{eff}}$ of the mode, effective refractive index $N$ of the mode, the wavelength and the refractive indices of the media. In the case of TM modes the fraction $\frac{n_c^2} {{n_m^2}}$ appears in the coupling coefficient (see Eqn. \eqref{eq:constant coupling coef application}). For molecular gratings, this fraction is essentially one because the refractive index of the cover medium $n_{{c}}$ is close to the refractive index of the perturbation $n_{{m}}$. For accurate treatment of background scattering the term $\frac{n_c^2} {{n_m^2}}$ should be retained, but has a rather small impact on the overall result compared to other uncertainties and measurement errors such as the determination of the correlation length \cite{Frutiger2019-um}.

Next, we consider three examples of diffractometric sensor configurations (Figure \ref{fig:optical configurations}b) for which we have calculated the coupling coefficients $\left|K\right|$ (see Table \ref{tab:coupling coefficients}). The first case represents a free space mode incident from the substrate and coupling to an outgoing mode leaving towards the substrate side (F$_s$/F$_s$ coupling). The second example involves an incident guided mode coupling to an outgoing free space mode leaving towards the substrate (G/F$_s$ coupling). For both cases, the azimuth $\varphi$ of the plane of incidence and the outgoing plane are chosen to be zero. The G/F$_s$ case was described and measured experimentally in previous papers \cite{Frutiger2019-um,Gatterdam2017-bk, Fattinger2014-pi}.\footnote{By choosing $\theta_{\text{out}}=0$, the formula stated here contains a factor ${{t_s^s\left( 0 \right)} \over {\sqrt {{n_s}} }}$, whereas the formula in \cite{Frutiger2019-um, Fattinger2014-pi} instead contains ${1 \over {\sqrt {{n_c}} }}$. The difference traces back to our use of out-modes, which fullfil the boundary conditions of Maxwell's equations. The previous equation \cite{Fattinger2014-pi} is based on results of Ref. \cite{Tamir1977-af} and the simplifying assumption that the influence of the interface is weak and that outcoupling in the substrate and cover is the same. Furthermore, the solution is only valid for perpendicular outgoing beams (no $\theta$-dependence). We can reproduce this solution if we neglect the interface in the out-mode and choose a uniform medium with refractive index $n_c$ for the out-arrangement. For $\theta_{\text{out}}=0$ the difference is marginal, as stated in Ref.\cite{Fattinger2014-pi}, but for other outcoupling angles and precise measurements the expression stated in this paper should be used.} The third configuration involves coupling between two guided modes with different in plane directions (G/G). For simplicity, we consider all modes to be TE (s) polarized to derive the same expression for TM (p) is straightforward. 

\begin{table}[h]
\def\arraystretch{2.5}
\label{tab:coupling coefficients}
\begin{tabular}{@{\hspace{0.5\tabcolsep}}c@{\hspace{0.5\tabcolsep}}|@{\hspace{0.5\tabcolsep}}c@{\hspace{0.5\tabcolsep}}|@{\hspace{0.5\tabcolsep}}c@{\hspace{0.5\tabcolsep}}l}
F$_s$F$_s$ &
 $\left|\frac{\pi}{\lambda n_s} \sqrt {{{1} \over {{A_b} A \cos\left(\theta_{\text{out}}\right)}}} t_s^s\left( {{\theta _{{\rm{in}}}}} \right) t_s^s\left( {{\theta _{{\rm{out}}}}} \right)\right|$ &
 1.56e13 $\frac{1}{\text{m}^3}$ \\
 GF$_s$ &
 $\left|{ \frac{\pi  t_s^s\left( {\theta _{{\rm{out}}}} \right)} {\lambda}}  \sqrt {{{\left( {n_f^2 - {N^2}} \right)} \over {N n_s \left( {n_f^2 - n_c^2} \right)}}{{2} \over {{w_b}{t_{{\rm{eff}}}}A \cos\left(\theta_{\text{out}}\right) }}}  \right|$ & 
4.64e14 $\frac{1}{\text{m}^3}$ \\
GG &
 $\left|{{2\pi } \over {\lambda N}}{{\left( {n_f^2 - {N^2}} \right)} \over {\left( {n_f^2 - n_c^2} \right)}}{{\cos \left( {2\varphi_{\rm{in}} } \right)} \over {{t_{{\rm{eff}}}}\sqrt {{w_b}w_{\rm{ch}}} }}\right|$ & 
4.81e15 $\frac{1}{\text{m}^3}$ 
 \\
 
\end{tabular}
\caption{Coupling coefficient $\left|{K}\right|$ for coupling between two free space modes (F$_s$/F$_s$), a guided mode and a free space mode (G/F$_s$), and two guided modes (G/G) for TE polarized incident and outgoing beams. The following numerical values were substituted: $\lambda$: 635e-9~m, $n_s$: 1.521, $n_c$: 1.33, $n_f$: 2.117, $N$: 1.814, $A_b$: 7.9e-7~m$^2$ (circular beam of diameter 1e-3~m), $A$: 1.26e-7~m$^2$ (equivalent to a circular molecular grating of diameter 4e-4~m), $w_b$: 1e-3~m, $t_{\rm{eff}}$: 329e-9~m, $\theta_{\rm{in}}$: 70$^\circ$, $\left| t_s^s\left( {{\theta _{{\rm{in}}}}}\right) \right|$: 1.41, $\theta_{\rm{out}}$: 0$^\circ$, $t_s^s\left( {{\theta _{{\rm{out}}}}} \right)$: 1.07 (In the GF$_s$ case, this is the Fresnel coefficient of the three layer interface, but the numerical value is the same. (we used a thickness for the waveguide of $t_f=145$e-9~m),  $\varphi_{\rm{in}}$ and  $\varphi_{\rm{out}}$ are 0$^\circ$ for the first two cases where a free space mode is involved. In the GG case, we assume the grating to have the shape of a parallelogram with one of its sides being parallel to the propagation direction of the outgoing beam. This assures that the outgoing beam is uniform of width corresponding to the height of the parallelogram. We arbitrarily choose $\varphi_{\rm{in}}=33.75^\circ$, $\varphi_{\rm{out}}=-33.75^\circ$. The characteristic width is simply the height of the parallelogram: $w_{\text{ch}}=a\sin(2\varphi)=$3.4e-4~m. For completeness, the TM polarization coefficients are given in Supplementary Information Section \ref{sec:TM_coupling_coefficients}}
\end{table}

Beams of guided modes are concentrated much more tightly to a waveguide underlying the sensing surface than freely propagating beams. Accordingly, for a given grating, the coupling coefficient $\left| {K} \right|$ is weakest for the coupling between freely propagating beams, stronger for coupling between a guided and a freely propagating beam and strongest for diffraction from a guided beam to a guided beam. This is reflected by the expressions for $\left| {K} \right|$ collected in Table \ref{tab:coupling coefficients}, which contain an additional small factor of order $t_{\rm{eff}}/w_b$ or $t_{\rm{eff}}/w_{\rm{ch}}$ for every freely propagating beam involved in the diffractive configuration. To show this quantitatively, we have evaluated the coupling coefficients of Table \ref{tab:coupling coefficients} for typical numerical values. Higher values of $\left| {K} \right|$ result in a stronger signal, \textit{i.e.} more diffracted photons per bound molecule. Quantitatively, this amounts to roughly two orders of magnitude stronger diffraction efficiency of GF vs FF and of GG vs GF coupling (Table \ref{tab:coupling coefficients} and Eqn. \eqref{eq:periodic_perturbation_evaluated}). However, it does not result in a better sensor performance as a larger $\left| {K} \right|$ also results in more background light.

\subsection*{Roughness induced signal to background}
Non-coherent scattering can limit the sensor performance as it causes background light to impinge on the detector. In order to achieve good sensor resolution it is usually required to maximize the signal to background ratio. As discussed above, often the dominating source for non-coherent background scattering is due to surface roughness. For a full assessment, also volume scattering from the substrate and the waveguide need to be taken into account \cite{Frutiger2019-um}. In addition, for many arrangements described in the literature, straylight from the source or the sample volume is limiting and not the surface roughness. This is mainly because of a non-ideal way of illuminating the coherent pattern (poor darkfield illumination). Nevertheless, for a well-designed diffractometric biosensor, the surface roughness often imposes a fundamental limitation to the sensor performance \cite{Frutiger2019-um}. For a background being dominated by surface roughness, the signal to background ratio derived from Eqns. \eqref{eq:periodic_perturbation_evaluated} and \eqref{eq:diffraction efficiency roughness} writes
\begin{equation}
\text{SBR} = {2 \over \pi }{{n_c^2} \over {{{\left( {n_d^2 - n_c^2} \right)}^2}}}{\left( {{{dn} \over {dc}}} \right)}^2{{{}\Gamma _{{\rm{coh}}}^2} \over {{\sigma ^2}}}{A \over {L_c^2}}{\left( {1 + \left|\vec\beta _g\right|^2 L_c^2} \right)^{3/2}}.
\end{equation}
This is a generalization of the figure of merit for focal molography defined in Eqn. (9) of Ref. \cite{Frutiger2019-um} (however the formulas in Ref. \cite{Frutiger2019-um} include roughness scattering from two interfaces). As one should expect, the signal to background ratio increases linearly with the sensor area, since the incoherent background is only proportional to the area, while the coherent diffraction scales with its square. It should be noted that the SBR is independent of the coupling coefficient $K$, and therefore of the optical configuration. The term $(1+|\vec{\beta_g}|^2L_c^2)^{3/2}/L_c^2$ has a minimum for $L_c= \sqrt{2}/|\vec{\beta_g}|$. Indeed, surface roughness impacts the SBR most unfavorably when its correlation length is of order of the inverse of the grating vector momentum $|\vec{\beta_g}|^{-1}$. The SBR could in principle be optimized by choosing the optical configuration with the largest possible $|\vec{\beta_g}|=\frac{2\pi}{\Lambda}$, \textit{i.e} the smallest grating vector. For diffraction between guided waves this favors a configuration close to normal incidence ($\varphi_{\text{in/out}}=0$) and Bragg reflection. In either case, the signal to background ratio still depends on the smoothness of the substrate and on the refractive index difference with respect to the cover. For a roughness dominated scattering background, different diffractometric arrangements will therefore have similar signal to background ratios.

\begin{table*}[ht]
    \centering
    \def\arraystretch{2.5}
\begin{tabular}{c|c|c|c|c|c|}
\multicolumn{2}{c|}{TE modes} & $E_{\rm{t}}\left(0\right)$ &$E_x\left(0\right)$ & $E_y\left(0\right)$ & $E_{\rm{n}}\left(0\right),E_z\left(0\right)$ \\ \hline
\multirow{2}{*}{F$_c$}  & in   &  $\sqrt {{{2{Z_0}P} \over {{n_c}{A_b}}}} \left( {1 + r_c^s\left( {{\theta _{{\rm{in}}}}} \right)} \right)$ & $ - {E_t}\sin \left( {{\varphi _{{\rm{in}}}}} \right)$ & ${E_t}\cos \left( {{\varphi _{{\rm{in}}}}} \right)$  & 0 \\ \cline{2-6} 
                   & out  & $\sqrt {{{2{Z_0}P} \over {{n_c}{A \cos\left(\theta_{\text{out}}\right)}}}} \left( {1 + r_c^s\left( {{\theta _{{\rm{out}}}}} \right)} \right)$ & $ - {E_t}\sin \left( {{\varphi _{{\rm{out}}}}} \right)$ & ${E_t}\cos \left( {{\varphi _{{\rm{out}}}}} \right)$ & 0 \\ \hline 
\multirow{2}{*}{F$_{s}$}  & in   & $\sqrt {{{2{Z_0}P} \over {{n_s}{A_b}}}} t_s^s\left( {{\theta _{{\rm{in}}}}} \right)$ & $- {E_t}\sin \left( {{\varphi _{{\rm{in}}}}} \right)$ & ${E_t}\cos \left( {{\varphi _{{\rm{in}}}}} \right)$  & 0 \\ \cline{2-6} 
                   & out  & $\sqrt {{{2{Z_0}P} \over {{n_s}{A \cos\left(\theta_{\text{out}}\right)}}}} t_s^s\left( {{\theta _{{\rm{out}}}}} \right)$ & $ - {E_t}\sin \left( {{\varphi _{{\rm{out}}}}} \right)$ & ${E_t}\cos \left( {{\varphi _{{\rm{out}}}}} \right)$  & 0 \\ \hline
\multirow{2}{*}{G}  & in   & $2\sqrt {{{\left( {n_f^2 - {N^2}} \right)} \over {N\left( {n_f^2 - n_c^2} \right)}}{{{Z_0}P} \over {{w_b}{t_{{\rm{eff}}}}}}}$ & $- {E_t}\sin \left( {{\varphi _{{\rm{in}}}}} \right)$ & ${E_t}\cos \left( {{\varphi _{{\rm{in}}}}} \right)$ & 0 \\ \cline{2-6} 
                   & out  & $2\sqrt {{{\left( {n_f^2 - {N^2}} \right)} \over {N\left( {n_f^2 - n_c^2} \right)}}{{{Z_0}P} \over {{w_{\rm{ch}}}{t_{{\rm{eff}}}}}}}$ & $ - {E_t}\sin \left( {{\varphi _{{\rm{out}}}}} \right)$ & ${E_t}\cos \left( {{\varphi _{{\rm{out}}}}} \right)$ & 0 \\ \hline

\multicolumn{2}{c|}{TM modes} & & & &  \\ \hline
\multirow{2}{*}{F$_c$}  & in  & $-\sqrt {{{2{Z_0}P} \over {{n_c}{A_b}}}} \left( {1 - r_c^p\left( {{\theta _{{\rm{in}}}}} \right)} \right)\cos \left( {{\theta _{{\rm{in}}}}} \right)$ & $ E_{\rm{t}}\cos \left( {{\varphi _{{\rm{in}}}}} \right)$ & $ E_{\rm{t}}\sin \left( {{\varphi _{{\rm{in}}}}} \right)$ &  $\sqrt {{{2{Z_0}P} \over {{n_c}{A_b}}}} \left( {1 + r_c^p\left( {{\theta _{{\rm{in}}}}} \right)} \right)\sin \left( {{\theta _{{\rm{in}}}}} \right)$\\ \cline{2-6} 
                   & out & $\sqrt {{{2{Z_0}P} \over {{n_c}{A}}}} \left( {1 - r_c^p\left( {{\theta _{{\rm{out}}}}} \right)} \right) \sqrt{\cos \left( {{\theta _{{\rm{out}}}}} \right)}$& $E_{\rm{t}}\cos \left( {{\varphi _{{\rm{out}}}}} \right)$ & $E_{\rm{t}}\sin \left( {{\varphi _{{\rm{out}}}}} \right)$ & $ - \sqrt {{{2{Z_0}P} \over {{n_c}{A }}}} \left( {1 + r_c^p\left( {{\theta _{{\rm{out}}}}} \right)} \right)\frac{\sin \left( {{\theta _{{\rm{out}}}}} \right)}{\sqrt{\cos\left(\theta_{\text{out}}\right)}}$ \\ \hline
\multirow{2}{*}{F$_{s}$}  & in & $\sqrt {{{2{Z_0}P} \over {{n_s}{A_b}}}} t_s^p\left( {{\theta _{{\rm{in}}}}} \right)\sqrt {1 - {{n_s^2} \over {n_c^2}}\sin^2 \left( {{\theta _{{\rm{in}}}}} \right)} $ & $E_{\rm{t}}\cos \left( {{\varphi _{{\rm{in}}}}} \right)$ & $E_{\rm{t}}\sin \left( {{\varphi _{{\rm{in}}}}} \right)$ & $-\sqrt {{{2{Z_0}P} \over {{n_s}{A_b}}}} t_s^p\left( {{\theta _{{\rm{in}}}}} \right){{{n_s}} \over {{n_c}}}\sin \left( {{\theta _{{\rm{in}}}}} \right)$ \\ \cline{2-6} 
                   & out & $- \sqrt {{{2{Z_0}P} \over {{n_s}{A}}}} t_s^p\left( {{\theta _{{\rm{out}}}}} \right)\frac{\sqrt {1 - {{n_s^2} \over {n_c^2}}\sin^2 \left( {{\theta _{{\rm{out}}}}} \right)}}{\sqrt{\cos\left(\theta_{\text{out}}\right)}} $& $E_{\rm{t}}\cos \left( {{\varphi _{{\rm{out}}}}} \right)$ & $ E_{\rm{t}}\sin \left( {{\varphi _{{\rm{out}}}}} \right)$ & $\sqrt {{{2{Z_0}P} \over {{n_s}{A}}}} t_s^p\left( {{\theta _{{\rm{out}}}}} \right){{{n_s}} \over {{n_c}}}\frac{\sin \left( {{\theta _{{\rm{out}}}}} \right)}{\sqrt{\cos\left(\theta_{\text{out}}\right)}}$ \\ \hline
\multirow{2}{*}{G}  & in & $i2\sqrt {{{\left( {n_f^2 - {N^2}} \right)} \over {N\left( {n_f^2 - n_c^2} \right){q_c}}}{{{Z_0}P} \over {{t_{{\rm{eff}}}}{w_b}}}} \sqrt {{{{N^2}} \over {n_c^2}} - 1} $ & $E_{\rm{t}}\cos \left( {{\varphi _{{\rm{in}}}}} \right)$ & $E_{\rm{t}}\sin \left( {{\varphi _{{\rm{in}}}}} \right)$ &  $-2\sqrt {{{\left( {n_f^2 - {N^2}} \right)} \over {N\left( {n_f^2 - n_c^2} \right){q_c}}}{{{Z_0}P} \over {{t_{{\rm{eff}}}}{w_b}}}} {N \over {{n_c}}}$\\ \cline{2-6} 
                   & out & $- i2\sqrt {{{\left( {n_f^2 - {N^2}} \right)} \over {N\left( {n_f^2 - n_c^2} \right){q_c}}}{{{Z_0}P} \over {{t_{{\rm{eff}}}}{w_{{\rm{ch}}}}}}} \sqrt {{{{N^2}} \over {n_c^2}} - 1} $ & $ E_{\rm{t}}\cos \left( {{\varphi _{{\rm{out}}}}} \right)$ & $E_{\rm{t}}\sin \left( {{\varphi _{{\rm{out}}}}} \right)$ & $2\sqrt {{{\left( {n_f^2 - {N^2}} \right)} \over {N\left( {n_f^2 - n_c^2} \right){q_c}}}{{{Z_0}P} \over {{t_{{\rm{eff}}}}{w_{{\rm{ch}}}}}}} {N \over {{n_c}}}$

\end{tabular}
\caption{Normalized field components at the grating location ($z=0^+$). They can be substituted to Eqn. \eqref{eq:constant coupling coef application}. This table contains all the expressions needed to calculated the coupling coefficient for any planar diffractometric sensor seamlessly (TE-TE, TE-TM, TM-TE, TM-TM coupling for any orientation and any combination of modes (F,G)). F$_c$ stands for a free space mode for a beam incident/outgoing from/to the cover. F$_s$ describes free space mode for a beam incident/outgoing from/to the substrate and G represents a guided mode of a dielectric slab waveguide. $A_b$ is the area of an incident free space beam and $w_b$ is the lateral width of an incident guided beam. Both are assumed to be larger than the molecular grating. $A$ is the area of the molecular grating. $w_{\text{ch}}$ is the linear size of the pattern (the characteristic width), as projected on a line orthogonal to the outgoing propagation direction. $\varphi$ is the azimuth angle taken from the positive grating normal and $\theta$ the out of plane angle with respect to the surface normal. The signs of the angles are defined in Figure \ref{fig:optical configurations}. The angle dependent Fresnel reflection $r$ and transmission coefficients $t$ are defined according to Ref. \cite{Novotny2011-pa,Born1999-tx} for the respective polarization. They can be defined for a two or three layer interface \cite{Frutiger2019-um, Novotny1997-hv}. If $\theta$ is larger than the critical angle total internal reflection occurs. It is to note that the transmitted angle becomes imaginary in this case $\left(\sqrt {1 - {{n_s^2} \over {n_c^2}}\sin^2\left( {{\theta _{{\rm{in}}}}} \right)}  \to i\sqrt {{{n_s^2} \over {n_c^2}}\sin^2 \left( {{\theta _{{\rm{in}}}}} \right) - 1} \right)$. $N$ is the effective refractive index of the guided mode and $t_{\rm{eff}}$ its effective thickness. ${q_c} = {\left( {{N / {{n_f}}}} \right)^2} + {\left( {{N / {{n_c}}}} \right)^2} - 1$ is a constant as defined in \cite{Kogelnik1975-xk}. No additional interfaces in the optical detection path (also not the backside of the chip) are taken into account in the formalism here. 
}
\label{tab:normalized_field_components}
\end{table*}

\subsection*{Characterization of diffractometric biosensors}

The theoretical tools of this paper are only useful if a diffractometric measurement is performed optimally. In the following, we suggest how to preferentially characterize a diffractometric sensor system and what key parameters should be stated to allow for comparisons. (For an example of a characterization see Ref. \cite{Frutiger2019-um}).

\begin{itemize}
    \item \textit{Measurement of the diffraction efficiency:} To compute the diffraction efficiency, two signals need to be measured accurately, the diffracted power and a reference power. The latter must allow computation of the incident power on the coherent pattern. The reference power is readily obtained from a measurement of the zeroth order beam, since the zeroth order is most often subject to the same reflection or transmission losses as the diffracted power. This allows straightforward calculation of the diffraction efficiency. For guided modes this can be accomplished by an outcoupling grating \cite{Frutiger2019-um}. The diffracted power can be obtained from a detector as described in Ref. \cite{Frutiger2019-um}. In addition, also the beam diameter should be measured and stated, because it is required for calculating the coupling coefficient.
    
    \item \textit{Background scattering measurement:} The mean background diffraction efficiency provides an estimate of the illumination quality and often determines the limit of detection in the case of an endpoint measurement \cite{Frutiger2019-um}. Even in real time experiments, minimizing the background is usually beneficial as this generally reduces the noise. Eqn. (\ref{eq:diffraction efficiency roughness}) helps with this analysis by providing an estimate for the background due to surface roughness. A measured value significantly above this estimated background indicates that the illumination configuration is not optimal. As a rule of thumb, a sensitive diffractometric biosensor has a background associated coherent mass density below 5 pg/mm$^2$.
    
    \item \textit{Measurement of the signal before target binding:} Many diffractometric sensors show significant diffraction before target binding (bias of the sensor) due to the diffraction at the immobilized receptors. A large bias contribution requires more precise intensity measurement. Therefore, the bias should be minimized if possible. This could be achieved by backfilling with a point mutated protein \cite{Fattinger2014-pi} or by covalently linked particles on the grooves. By measuring $\Gamma_{\rm{coh}}$ before the target binding the resolution associated with the bias can be determined.
    
    \item \textit{Determination of sensor resolution:} The resolution of a sensor can be computed from the root mean square (rms) mass baseline noise. We suggest to measure the baseline noise of a functionalized sensor in air, in degased buffer and in an appropriate biofluid at zero target concentration. This measurement should be performed over the time scale of a real experiment (ca. 20 min) without any drift correction. If a proper noise characterization is required, the power spectral density should be computed and stated \cite{Leuermann2019-fk}. State of the art diffractometric biosensors have resolutions of $\Gamma_{\rm{coh}}$ below 100 fg/mm$^2$ over 20 min \cite{Frutiger2019-um}.
\end{itemize}

\section*{Conclusion}
Diffraction based biosensors have great potential in overcoming current limitations of biomolecular interaction analysis in complex environments. In addition, they require much less sensor equilibration time than refractometric sensors and they are insensitive to temperature and buffer changes. So far, the diffractometric biosensing field has suffered from the lack of a universal parameter to quantify the sensor output. This shortcoming prevented the efficient comparison among different diffractometric sensor arrangements and limited their comparison to the established refractometric technologies. The coherent surface mass density $\Gamma_{\rm{coh}}$ is an unifying quantitative parameter as it is independent of the assay, the type of diffraction grating or the sensor arrangement. In addition, $\Gamma_{\rm{coh}}$ can be converted to the total surface mass density, $\Gamma_{\rm{tot}}$, which can be used for comparison to other sensor technologies. To compute $\Gamma_{\rm{coh}}$ with the described formulae only the measurement of the diffraction efficiency and the beam diameter is required. Besides being useful for mass quantification, these formulae can also be applied to study non-coherent effects such as surface roughness or non-specific binding. Especially, in the case of surface roughness, these effects may limit the sensor performance. Careful characterization of the sensor configuration can lead to a tremendous improvement in resolution. We believe that the tools we provided here to quantify, characterize and compare diffractometric biosensors will lead to significant advancement of the field and help to establish a reliable alternative to refractometric biosensors, especially in applications involving complex biofluids.

\begin{acknowledgments}
Janos Vörös is acknowledged for valuable input and discussions during the entire project. 
\end{acknowledgments}

\bibliographystyle{IEEEtran}
%bibliographystyle{apa} %for alphabetical
\bibliography{references}

\appendix

\section{Author contributions}

Y.B. and A.F. derived the analytical expressions with critical input and guidance from M.M.. M.M. assured that the nomenclature and assumptions are physically rigorous. Y.B. and A.F. made the figures. C.F., A.F. and Y.B. planned the content of the work and developed the terminology for quantitative diffractometric biosensing. Y.B., A.F., M.M. and C.F. wrote the manuscript. R.D. derived the initial expression from coupled mode theory for coupling between two guided modes. A.M.R assured that the main results and conclusions are in a format that can be used and understood by experimentalists. All authors have read and approved the manuscript for submission.

\end{document}